\documentstyle[12pt]{article}
\date{}

\def\lsim{\buildrel {\textstyle <}\over {_\sim}}
\begin{document}
\baselineskip 8mm
\sloppy
\thispagestyle{empty}
%\maketitle
\begin{center}
{\Large\bf  Chiral Glass: \ A New Equilibrium Phase} 
\end{center}
\begin{center}
{\Large\bf  of Ceramic High-$T_c$ Superconductors}
\end{center}

\vspace*{1cm}

\begin{center}
Hikaru Kawamura 

{\sl Faculty of Engineering and Design, 
Kyoto Institute of Technology, 

Sakyo-ku, Kyoto 606, Japan}
\end{center}

\begin{center}
Mai Suan Li  

{\sl Institute of Physics, Polish Academy
of Sciences,  02-668 Warsaw, Poland}
\end{center}
\vspace*{1cm}

\noindent
{\bf Abstract}

Possible occurrence of an equilibrium thermodynamic phase with
a spontaneously broken time-reversal symmetry is studied in a 
model ceramic superconductor with anisotropic pairing symmetry.
It is shown by Monte Carlo simulations that such a ``chiral-glass''
phase is truly stable even under the influence of screening.
Existence of frustration in zero external field, 
arising from the $d\/$-wave pairing symmetry of high-$T_c$
superconductors, is essential to realize this phase.
Via a finite-size scaling analysis, critical exponents associated 
with the chiral-glass transition are estimated to be 
$\nu _{{\rm CG}}=1.3\pm 0.2$ and $\eta _{{\rm CG}}=-0.2\pm 0.2$.
These values turn out to be close to those of the Ising spin glass.
Phase diagram of the model is constructed and the
implications to experiments are discussed.

\begin{flushright}
e-mail: kawamura@hiei.kit.ac.jp
\end{flushright}


\hfill

\newpage
\noindent
\section {Introduction}
Among a variety of macroscopic thermodynamic properties 
of superconductors, the type and
the nature of possible thermodynamic phases is
of central importance. 
For example, considerable attention has recently been paid to the 
possible phase of high-$T_c$ superconductors
in applied magnetic fields.$^1$ 
For sufficiently random or
dirty samples,
the existence of
a true thermodynamic phase with zero linear resistance
was predicted (a vortex-glass phase).$^2$
Recent numerical works suggest, however,  that the screening effects
eventually destabilize the vortex-glass phase.$^3$
In zero external field, by contrast, 
the only thermodynamic phase known to date either in clean or
random superconductors is the standard Meissner phase.

Meanwhile, recent experimental studies have revealed that cuprate
high-$T_c$ superconductors have an anisotropic pairing
symmetry, probably of the $d_{x^2-y^2}$-wave type.$^4$
Naturally, one may expect that such anisotropic nature of the
superconducting order parameter could give rise to  
novel  thermodynamic properties
not encountered in the conventional $s$-wave superconductors,
possibly the  appearance of a new thermodynamic phase.
Unfortunately, this appears not to be the case
in clean single crystals, although enhanced effects of thermal
fluctuations give rise to a variety of interesting phenomena
in high-$T_c$ superconductors. 
This is because
the $d_{x^2-y^2}$-wave order parameter
is characterized by a single phase variable of the 
condensate as in the 
conventional superconductors, and the resulting
Ginzburg-Landau Hamiltonian, which is known to well describe 
various macroscopic properties of superconductors, has
essentially the same form
as in the conventional superconductors.
By contrast, in {\em ceramic\/} or {\em granular\/} high-$T_c$ 
samples,
the situation may well
differ because 
ceramic samples can be regarded as a random Josephson network
and the anisotropic superconducting
order parameter largely modifies 
the properties of the Josephson junction.

One remarkable effect is the appearance of the `$\pi $ junction'
characterized by the {\em negative\/} Josephson coupling, $J<0$,
across which the order parameter changes the phase by $\pi $.
One important consequence of such $\pi $ junctions is the
appearance of frustration even 
in zero external field.$^5$  Frustration is
borne by the `odd ring', a closed junction-loop having 
odd numbers of $\pi $ junctions. 
On the basis of a single-loop model in which
high-$T_c$ ceramics were regarded as an ensemble of noninteracting 
junction-loops, Sigrist and Rice$^6$ successfully
explained the paramagnetic Meissner effect
observed  experimentally in certain high-$T_c$ ceramic 
samples.$^{7,8}$
Magnetic moments spontaneously induced   at the
odd rings give rise to a paramagnetic
response observed experimentally.

It is also evident, however,  
that the paramagnetic  Meissner effect itself
is  not directly related
to a new thermodynamic phase,
since an ensemble of noninteracting junction-loops is enough
to cause the paramagnetic susceptibility,$^{6,9}$
just as an ensemble of free spins gives rise to the paramagnetic
Curie law. 
In the present paper, 
we wish to address the question
whether a new type of equilibrium phase closely related to the
unconventional pairing symemtry of high-$T_c$ superconductors
is ever possible
in certain ceramic samples.
Indeed, one of the present authors (H.K.) recently proposed that
such a novel thermodynamic phase might be realized
in zero external field
in certain ceramic high-$T_c$ superconductors.$^{10}$
This state is characterized by a spontaneously
broken time-reversal symmetry with keeping the $U(1)$
gauge symmetry, and is called a `chiral-glass phase'.$^{11}$
The order parameter is then a `chirality',
which represents the direction of
the local loop-supercurrent over grains. From a 
symmetry consideration, 
the nonlinear susceptibility was predicted to
diverge with a negative sign at the associated
chiral-glass transition point. 
Frustration effect, which arises due to the random
distribution of
$\pi $ junctions with the negative Josephson coupling,  
is essential to
realize this phase. Note that in this chiral-glass state, 
unlike in the 
vortex-glass (gauge-glass) state under external fields, 
the phase of the condensate
is {\em not\/} ordered, even randomly, 
on sufficient long length and time scales: 
The ordering occurs only in the loop-supercurrents, or in the
{\it chiralities\/}. 

There are several experimental results 
which appear to corroborate
the existence of such a novel glassy zero-field phase in
ceramic high-$T_c$ superconductors.
Since the  discovery of
high-$T_c$ superconductors, it has been known that 
ceramic high-$T_c$ samples often
exhibit a glassy behavior reminiscent of 
the spin glass.$^{12-14}$
%in week or even in zero external field. 
More recently, 
Leylekian, Ocio and Hammann observed 
via the noise and ac susceptibility measurements that LSCO ceramic
samples showed a glassy behavior 
reminiscent of the spin glass {\em even in zero 
external field\/}.$^{15}$ 
They also observed an intergranular 
cooperative phenomenon indicative of a glassy
phase transition. It appears natural to interpret this cooperative
phenomenon in terms of the 
proposed chiral-glass picture.
More direct support of the chiral glass  has recently been
reported by the ac susceptibility measurements on 
YBCO ceramic samples. 
Thus, Matsuura {\em et al\/} observed a negatively divergent  
nonlinear susceptibility 
at an intergranular transition point,$^{16}$
consistent with the proposed chiral-glass picture.

It should be emphasized here,
however, that the theoretical analysis of Ref.10 was based on an
analogy to the {\em XY\/} spin glass,$^{17}$ and completely
neglected the effects of screening  (coupling of the condensate to
fluctuating magnetic fields). 
Thus, the fate of the proposed chiral-glass phase in the presence
of screening is not yet clear. 
It should  be  noted that the screening effect could be
substantial in the intergranular ordering of ceramic high-$T_c$ 
materials,
since the length unit to be compared with the penetration
depth is 
the grain size ($\sim 1\mu m$) 
rather than the short coherence length of the Cooper pair.
As the screening effect makes 
the otherwise long-ranged interaction
between the chirality (quenched-in half a vortex) short-ranged, 
one may wonder if it would
eventually wash out a sharp phase transition
and destabilize the chiral glass phase, just as it destabilizes
the vortex-glass phase of type-II superconductors in a field.

The purpose of the present paper is to investigate the 
stability of the  
hypothetical chiral-glass phase against this screening effect,
and to determine whether the chiral-glass state
really exists as a true thermodynamic phase or not.
For this purpose, 
we perform extensive Monte Carlo simulations
on a frustrated 
three-dimensional lattice model introduced 
by Dom\'inguez {\em et al\/},$^{18}$ 
in which the
phase variables located at each site of the lattice are coupled 
to the fluctuating magnetic-field variables at each link.
The model can also viewed as a spin-glass-type ({\em i.e., }
random and frustrated) version  of a $U(1)$ lattice gauge theory.
The same model was studied previously by Monte Carlo simulations
with an interest in the behavior of linear$^{18}$ and nonlinear$^{19}$
susceptibilities.
But, neither of these previous simulations was fully 
equilibrated,
and thus, could not give any information 
whether a true equilibrium 
phase could exist or not. In the present paper, we perform
equilibrium simulations  on the same model based on an
extended ensemble method recently proposed 
by Hukushima and Nemoto,$^{20}$ trying to determine the equilibrium
properties of the model.
A part of  the results was already reported in Ref.21.
\par
The remainder of this paper is organized as follows.
In \S 2, we introduce our model and several physical quantities
of  interest.  In \S 3, we give technical 
details of our Monte Carlo simulation.
The results of Monte Carlo simulation
are presented and analyzed in \S 4. 
By studying the Binder ratio
associated with the chirality, we show
that there 
indeed exists a stable chiral-glass phase
with a spontaneously broken time-reversal symmetry 
even in the presence of screening.
Critical exponents characterizing the chiral-glass
transition are determined with use of a finite-size scaling analysis.
Phase diagram of the model is also constructed. 
Section 5 is devoted to summary and discussion. 
We discuss in some detail the
implications of the obtained results 
to the possible experimental detection of the chiral-glass state
in ceramic high-$T_c$ superconductors. In particular, requirements 
for the appropriate samples as well as the method of detection are 
examined.
\par
\bigskip\bigskip
\noindent
\setcounter{equation}{0}
\renewcommand{\theequation}{2.\arabic{equation}}
\section {Model}
We assume that weak links 
connecting the neighboring grains are distributed 
sufficiently dense, so that the system can be viewed as an
infinite network of Josephson junctions which are 
not decomposed into finite clusters.  Putting  superconducting
grains at the sites of a simple-cubic lattice, 
we model  such ceramic superconductors by 
a three-dimensional lattice model
of Josephson-junction array with finite self-inductance.
Neglecting the charging effects of the grain,
we consider the following zero-field Hamiltonian,$^{18,19}$
\begin{eqnarray}
{\cal H} = - \sum _{<ij>} J_{ij}\cos (\theta _i-\theta _j-A_{ij})
+ \frac {1}{2{\cal L}} 
(\frac {\phi _0}{2\pi })^2 \sum _p (\vec \nabla \times \vec A)^2, 
\end{eqnarray}
where $\theta _i$ is the phase of the condensate of the grain
at the $i$-th site of a simple cubic lattice,
$\vec A$ being the fluctuating gauge potential at each link
of the lattice,
$\phi _0$ the flux quantum,
$J_{ij}$ the Josephson coupling
between the $i$-th and $j$-th grains, 
and the lattice curl $\vec \nabla \times \vec A$
is the directed sum of $A_{ij}$'s round
a plaquette.
${\cal L}$ is the self-inductance of a loop (an elementary plaquette),
while the mutual inductance between different loops
is neglected.
The first sum is taken over all nearest-neighbor pairs, while the
second sum is taken over all elementary plaquettes on the lattice.
Fluctuating  thermodynamic
variables to be summed over are the phase variables,
$\theta _i$, at each site and the gauge variables, $A_{ij}$, at each
link.

In this model, quenched randomness occurs
only in the distribution of the Josephson coupling, $J_{ij}$, 
which is assumed to
be an independent
random variable taking the values 
$J$ or $-J$ with equal probability ($\pm J$ or binary 
distribution), each
representing $0$ and $\pi $  junctions.
We also assume that
$J_{ij}$ is 
independent of temperature and magnetic field. This assumption is more
or less justified when the  intergranular ordering occurs
at a temperature  considerably lower than the 
superconducting transition temperature of the grain.
While our simulation is performed 
for this particular distribution of
$J_{ij}$, one could expect from experience in spin-glass studies
that the results would be 
rather insensitive to the details of the
distribution, {\em e.g.\/}, 
a slight asymmetry between $\pm J$ or the detailed 
form of the distribution.
\par
The bare Josephson penetration depth  
in units of lattice spacing is
given by 
\begin{eqnarray}
\lambda _0=1/\sqrt{\tilde {\cal L}},
\end{eqnarray}
where $\tilde {\cal L}$ is the dimensionless inductance defined by
\begin{eqnarray}
\tilde {\cal L}=(2\pi /\phi_0)^2 J{\cal L}.
\end{eqnarray}
Thus, larger inductance  corresponds to smaller
penetration depth with enhanced effects of screening.
In the limit $\tilde {\cal L}\rightarrow 0$, or
$\lambda _0\rightarrow \infty $, the screening effect becomes negligible
and one recovers the $XY$ spin-glass Hamiltonian of Ref.10.
In the opposite limit $\tilde {\cal L}\rightarrow \infty $, on the
other hand, it can be shown that
the model reduces to the noninteracting one.$^{19}$ Therefore,
the system remains in the disordered state 
even at $T=0$ in this limit.

The symmetry property of the Hamiltonian (2.1) was analyzed
in detail in Ref.19.
Contrary to the vortex-glass (gauge-glass)
Hamiltonian, the Hamiltonian (2.1), defined in zero field,
keeps the $Z_2$ time-reversal symmetry in addition to the $U(1)$
gauge symmetry. 
Frustration arises  from the random
distribution of both positive and negative Josephson couplings. 
This should be contrasted to
the vortex-glass (gauge-glass) problem, where
the  associated Hamiltonian does not possess
the time-reversal symmetry 
due to external magnetic fields, while  
the frustration arises from  the magnetic field, 
not from the $J_{ij}$.
\par
The Hamiltonian (2.1) can also be written in the dimensionless form as
\begin{eqnarray}
\tilde{\cal H} \equiv \frac {{\cal H}}{J}
= - \sum _{<ij>} \tilde J_{ij}\cos (\theta _i-\theta _j-A_{ij})
+ \frac {1}{2\tilde {\cal L}} 
\sum _p (\vec \nabla \times \vec A)^2,
\end{eqnarray}
where $\tilde J_{ij}\equiv J_{ij}/J$ is the dimensionless
quenched random variable, 
taking values $\pm 1$ with equal probability.

The local chirality may be 
defined at each plaquette by the gauge-invariant
quantity,$^{19}$
\begin{eqnarray}
\kappa_p=2^{-3/2}\sum_{<ij>}^p 
\tilde J_{ij}\sin (\theta_i-\theta_j-A_{ij}),
\end{eqnarray}
where the sum runs over a directed contour
along the sides of the plaquette $p$. 
Physically, the chirality,
$\kappa _p$, 
is a half ($\pi $) vortex, being proportional to the  
loop-supercurrent circulating round a  plaquette $p$.
If the plaquette $p$ is frustrated, the local chirality
$\kappa _p$ tends to take a finite value, its sign
representing either clockwise or counterclockwise orientation of
circulating
supercurrent. If, on the other hand, 
the plaquette is unfrustrated, 
it tends to take a value around zero. 
Note that the chirality is a pseudoscalar in the sense that it is
invariant under global $U(1)$ gauge transformation, 
$\theta _i\rightarrow \theta _i+\Delta \theta,\ A_{ij}\rightarrow
A_{ij}$,
but changes its sign under global $Z_2$ time-reversal transformation,
$\theta _i\rightarrow -\theta _i,\ A_{ij}\rightarrow
-A_{ij}$.
Due to this symmetry property, 
chirality can be regarded
as an order
parameter of the chiral order.

Induced local flux or magnetization
threading a plaquette $p$ is given in the
dimensionless form by, 
\begin{eqnarray}
f_p=\frac {\Phi_p}{\phi_0},\ \ \ \ \  \Phi_p=\vec \nabla 
\times \vec A. 
\end{eqnarray}
Flux is also a pseudoscalar like  chirality, whose
sign represents the direction of the  induced
magnetic moment threading the plaquette $p$.
Total magnetization per plaquette along the $z\/$-axis
is given by 
\begin{eqnarray}
m=\frac {1}{4\pi SN_p}\sum_{p\in <xy>}
\Phi _p,
\end{eqnarray}
where $S$ is the area of a plaquette and
the sum is taken over all $N_p$ 
plaquettes on the $<xy>$ plane of the lattice.
The corresponding dimensionless quantity, $\tilde m$, 
can be defined by
\begin{eqnarray}
\tilde m\equiv \frac {4\pi S}{\phi_0}m
=\frac {1}{N_p}\sum_{p\in <xy>}f_p.
\end{eqnarray}

The linear susceptibility, $\chi $, is given by$^{18,19}$
\begin{eqnarray}
\chi \equiv \frac {{\rm d}m}{{\rm d}H}
=\frac {\pi \tilde \beta N_p}{\tilde {\cal L}}
[<\tilde m^2>]-\frac {1}{4\pi},
\end{eqnarray}
where $H$ is an external magnetic field, $\tilde \beta$ is the
dimensionless inverse temperature defined by
$\tilde \beta=J/k_BT$,
$<\cdots >$ represents 
a thermal average, and  $[\cdots ]$  represents 
a configurational average over the bond distribution.
The linear susceptibility is dimensionless in cgs units.
The nonlinear susceptibility, $\chi _2$, or its dimensionless
counterpart, $\tilde \chi _2$, is given by$^{19}$
\begin{eqnarray}
\chi_2=\frac {1}{6}\frac {{\rm d}^3 m}{{\rm d}H^3}
\equiv (\frac {4\pi S}{\phi_0 })^2\tilde \chi_2 ,
\cr
\tilde \chi_2 =
\frac {1}{6}(\frac {\pi  \tilde \beta N_p} {\tilde {\cal L}})^3
[<\tilde m^4>-3<\tilde m^2>^2].
\end{eqnarray}
In these expressions of $\chi $ and $\chi _2$, we omit
the parts which are odd under the
time-reversal operation, $m\rightarrow -m$:
In zero external field, 
such odd parts should vanish identically  in any finite
system in full equilibrium.
Note that the above $\chi _{2}$,  being
proportional to the minus of the third-harmonic
component of the ac susceptibility, is sometimes denoted as $\chi_3$
in the literature.
\par
\bigskip
\bigskip

\setcounter{equation}{0}
\renewcommand{\theequation}{3.\arabic{equation}}
\noindent
\section {Monte Carlo simulation}
We perform  Monte Carlo simulations for the Hamiltonian (2.1)
based on the standard Metropolis method combined with an extended
ensemble method.
Several values of the 
dimensionless inductance, $\tilde {\cal L}$, are simulated 
including $\tilde {\cal L}=1,3,4,5$, with the greatest 
effort in the case of
$\tilde {\cal L}=1$. Since
the bare penetration depth,  
$\lambda _0$, in those cases 
is equal to, or less than one lattice spacing,
we expect that the effect of screening should  manifest
itself even for  rather small lattices studied here, which contain
$L\times L\times L$  sites with $L=3,4,6,8,10$.
Sample average is taken over 1540 ($L=3$),  1000 ($L=4$),
500 ($L=6$), 300 ($L=8$) and 100 ($L=10$) independent
bond realizations.

We impose free boundary conditions  on all sides of the lattice.
If, on the other hand, 
one applies the standard periodic boundary conditions
on the link variables $A_{ij}$'s, one has somewhat 
unphysical results that the magnetization 
vanishes trivially even under external fields. 
In zero field, the periodic boundary conditions imposed on the
link variables $A_{ij}$'s  
also lead to the trivial vanishing of the susceptibility. 
In view of such unphysical nature of the periodic boundary
conditions, we impose free boundary
conditions in the following simulations. 

Monte Carlo simulation is performed
according to the version of an extended ensemble method of
Ref.20,
where the whole
configurations at two neighboring temperatures of the same sample
are occasionally exchanged. 
In this method, one simulates the sample 
with a given bond realization 
at $N_T$ distinct temperatures at a time
distributed in the range [$T_{min}, T_{max}$].
Monte Carlo updating 
consists of the two parts: The first part is the standard 
local Metropolis updating
at each temperature, and the second part is an exchange of the whole 
lattices at two neighboring temperatures.

Since the present spin-glass-like model possesses the link variable,
$A_{ij}$, 
in addition to the site variable, $\theta _i$, 
an equilibrium simulation
is rather hard even with the new efficient algorithm.
In the case of $L=8$, for example, we prepare 20
temperature points distributed in the range [0.27J, 0.8J] 
for a given sample, 
and perform $1.5\times 10^5$
exchanges  per temperature of the whole lattices
combined with the same number of standard 
`single-spin-flip' Metropolis sweeps.$^{20}$
Equilibration is checked by monitoring the 
stability of the results
against at least three-times longer runs for a subset of samples.

As long as  one is interested in the gauge-invariant quantities
like the chirality or the flux, the results would not depend on the 
particular choice of the gauge.
In most of our calculation,
we choose the gauge where 
the $A_{ij}$'s along the $z$-direction are fixed to
be zero.  We also use other gauges to take some 
limited data, including 
the Coulomb gauge and the `temporal gauge',$^{18,19}$ 
just to make sure that the results
are really independent of the choice of the gauge.

We run  in parallel two independent replicas 
with the same bond realization
and compute an overlap between the chiral variables in the two
replicas,$^{17}$
\begin{eqnarray}
q_{\kappa } =\frac {1}{N_p}\sum _p \kappa _p ^{(1)}
\kappa _p ^{(2)}. 
\end{eqnarray}
In terms of this chiral overlap, $q_{\kappa} $, 
the Binder ratio of the chirality is calculated by
\begin{eqnarray}
g_{{\rm CG}}=\frac {1}{2}
(3-\frac {[<q_\kappa ^4>]}{[<q_\kappa ^2>]^2}). 
\end{eqnarray}
Here $g_{{\rm CG}}$ is normalized 
so that in the thermodynamic limit 
it tends to zero above the chiral-glass
transition temperature, $T_{{\rm CG}}$, and
tends to unity below $T_{{\rm CG}}$ provided
the ground state is non-degenerate. At the chiral-glass
transition point,
curves of $g_{{\rm CG}}$ against $T$ 
for different $L$ should intersect asymptotically. 

The chiral-glass susceptibility, which is expected to diverge at the
chiral-glass transition point, is given by
\begin{eqnarray}
\chi _{{\rm CG}}=N_p[<q_\kappa ^2>],
\end{eqnarray}
The behavior of the chiral short-range order may be seen
via the mean magnitude of the local chirality, $\bar \kappa $, 
defined by
\begin{eqnarray}
\bar \kappa =\{\frac {1}{N_p}\sum _p [<\kappa_p^2>]\}^{1/2}.
\end{eqnarray}
Note that for the frustrated noncollinear configurations,
$\bar \kappa $, tends to take a finite value whereas
for the unfrustrated collinear configurations
$\bar \kappa $ tends to vanish.
One can also define a reduced chiral-glass
susceptibility, $\tilde \chi _{{\rm CG}}$, 
corrected for the short-range order effect by
dividing $\chi _{{\rm CG}}$ 
by the appropriate power of the magnitude of the
local chirality,$^{17}$
\begin{eqnarray}
\tilde \chi_{{\rm CG}} =\chi _{{\rm CG}} /\bar \kappa^4.
\end{eqnarray}

Similarly to the case of the chirality, one can introduce 
an overlap 
between the flux variables in the two
independent replicas,
\begin{eqnarray}
q_{f} =\frac {1}{N_p}\sum _p f _p ^{(1)}
f _p ^{(2)}. 
\end{eqnarray}
In terms of $q_{f} $, 
the Binder ratio of the flux is calculated by
\begin{eqnarray}
g_{{\rm FG}}=\frac {1}{2}(3-\frac {[<q_f ^4>]}{[<q_f ^2>]^2}).
\end{eqnarray}
The flux-glass susceptibility, $\chi _{{\rm FG}}$, 
and its reduced counterpart, 
$\tilde \chi _{{\rm FG}}$, are defined by
\begin{eqnarray}
\chi _{{\rm FG}}=N_p[<q_f ^2>],
\\
\tilde \chi_{{\rm FG}} =\chi _{{\rm FG}} /\bar f^4,
\end{eqnarray}
respectively, where the
the mean magnitude of the local flux is defined by
\begin{eqnarray}
\bar f =\{\frac {1}{N_p}\sum _p [<f_p^2>]\}^{1/2}.
\end{eqnarray}
\bigskip
\bigskip

\setcounter{equation}{0}
\renewcommand{\theequation}{4.\arabic{equation}}
\noindent
\section {Monte Carlo results}
\smallskip\par
In this section, we present the results of our Monte Carlo 
simulations.
Most extensive simulation is made for the inductance
$\tilde {\cal L}=1$, which corresponds to the bare penetration depth,
$\lambda _0$,
equal to one lattice spacing. In the subsection [A], we present
our Monte Carlo 
results for this inductance,
$\tilde {\cal L}=1$. The results for other inductactances
including $\tilde {\cal L}=3,4,5$ will be presented later in the 
subsection [B].
\par\medskip
\noindent
[A]\  $\tilde {\cal L}=1$
\par
In Fig.1, the temperature dependence of 
the root-mean square of the local-chirality amplitude, $\bar \kappa $,
defined by Eq.(3.4), is shown for
various lattice sizes. 
Even at lower temperatures $\bar \kappa $ keeps a nonzero value, 
slightly increasing with decreasing temperature, which 
indicates that the
chiral short-range order is 
developed in  the temperature range studied here. 
 
Fig.2a displays the size and temperature dependence of the
Binder ratio of the chirality, 
$g_{{\rm CG}}$.
The data of $g_{{\rm CG}}$ for 
$L=3,4,6,8$ all cross at almost the same
temperature $T\sim 0.28-0.29$, strongly
suggesting the occurrence of a  finite-temperature
chiral-glass transition at $T_{{\rm CG}}=0.286\pm 0.01$ (temperature $T$ is
measured in units of $J$). In particular, 
the data below $T_{{\rm CG}}$ show
a rather clear fan out.

The determined  value of $T_{{\rm CG}}$ is slightly lower than the
corresponding chiral-glass transition temperature of the 
pure $\pm J$ {\em XY\/} spin glass 
determined in Ref.17, $T_{{\rm CG}}=0.32\pm 0.01$.
Note that 
the spin-glass model corresponds to the $\tilde {\cal L}\rightarrow 0$
limit of the present model.
The observed suppression of $T_{{\rm CG}}$ by the screening effect 
seems reasonable, 
since the screening effect
makes the long-ranged interaction between vortices
(chiralities) short-ranged, which should make 
the  chiral-glass transition less favorable.

The value of the Binder ratio at the transition point, 
$g^*_{{\rm CG}}$,
is estimated to be $g^*_{{\rm CG}}\simeq 0.38$. 
The estimated value is
considerably smaller than the corresponding value
of the 3D {\em XY\/}
spin glass,$^{17}$ 
$g^*_{{\rm CG}}\simeq 0.72$. This large deviation probably comes from
the difference in the choice of boundary conditions, {\em i.e.\/},
free boundary
conditions in the present simulation and periodic boundary
conditions in Ref.17. Note that the value of 
$g^*$ is known to depend on
the choice of boundary conditions even in a given universality class.

Standard finite-size scaling analysis is made for $g_{{\rm CG}}$
based on the one-parameter fit of the form,
\begin{eqnarray}
g_{{\rm CG}}=\bar g_{{\rm CG}}
(L^{1/\nu _{{\rm CG}}}\mid T-T_{{\rm CG}}\mid ),
\end{eqnarray}
with fixing $T_{{\rm CG}}=0.286$, where $\bar g_{{\rm CG}}$ is
a scaling function. Then, 
the chiral correlation-length exponent $\nu _{{\rm CG}}$ is estimated
to be $\nu _{{\rm CG}}=1.3\pm 0.2$. The corresponding
finite-size-scaling plot in given in Fig.2b.

The temperature and size dependence of 
the reduced chiral-glass
susceptibility, $\tilde \chi _{{\rm CG}}$, defined by eq.(3.5), 
are shown in Fig.3a.
Finite-size scaling analysis based on the relation,
\begin{eqnarray}
\tilde \chi _{{\rm CG}}=L^{2-\eta _{{\rm CG}}}
\overline {\tilde \chi }_{{\rm CG}}
(L^{1/\nu _{{\rm CG}}}\mid T-T_{{\rm CG}}
\mid ),
\end{eqnarray}
is made with fixing $T_{{\rm CG}}=0.286$ and $\nu _{{\rm CG}}=1.3$, 
yielding the chiral critical-point-decay
exponent $\eta _{{\rm CG}}=-0.2\pm 0.2$. The resulting 
finite-size-scaling plot is displayed in Fig.3b.
Other exponents can be estimated via 
the standard scaling relations
as
$\gamma _{{\rm CG}}\simeq 2.9$ and $\beta _{{\rm CG}}\simeq 0.5$.

The obtained chiral-glass 
exponents are rather 
close to the values determined previously for the $\pm J$ 
{\em XY\/} spin glass, {\em i.e.\/}, the model without
screening; $\nu _{{\rm CG}}=1.5\pm 0.3$ and 
$\eta _{{\rm CG}}=-0.4\pm 0.2$.$^{17}$
Therefore, our present result seems consistent with the 
view that the screening effect is actually 
irrelevant at the 3D chiral-glass transition.
%although in view of the associated large error bars we cannot give a
%truly definitive conclusion.

It should be noted here that
the determined chiral-glass exponents are also 
not far from the standard 
spin-glass exponents of the 3D Ising
spin glass.$^{22,23}$ In recent Monte Carlo simulations of
the 3D Ising
spin glass, 
however, 
considerably different values were reported for the exponent $\nu $,
depending on whether
$\nu $ was determined from the scaling of the
Binder ratio, $g$, or from the scaling of the
spin-glass 
susceptibility, $\chi _{{\rm SG}}$. 
In Ref.22, for example, 
the former procedure  gave an estimate $\nu \simeq 2.0$ while the 
latter procedure gave $\nu \simeq 1.6$,
whereas in Ref.23, the former  gave $\nu \simeq 3.0$ and the 
latter  gave
$\nu \simeq 1.5$. By contrast, in the present simulation, 
we did not observe such significant
discrepancy between the estimated values of $\nu _{{\rm CG}}$:
A common value of $\nu _{{\rm CG}}\simeq 1.3$ gave  
satisfactory fits both for $g_{{\rm CG}}$ and
$\tilde \chi _{{\rm CG}}$. 
At present,
we are not sure whether such apparent difference from 
the standard Ising spin glass is simply due to 
finite-size effects, or is suggesting 
the possibility that the universality class
of the chiral-glass transition and that of
the Ising spin glass  are in fact different. 
Since the chirality can be viewed as 
a ``continuous'' Ising
variable,
there exists an obvious similarity between
the chiral glass and the pure Ising
spin glass from a symmetry viewpoint.  
%with the frustrating interaction working between them.
By contrast, in the present model, 
there exists a local constraint
on the possible configurations of the chiral variables,$^{24}$ 
which is
absent in the Ising spin glass. This  may possibly change
the universality class of the transition.
Further studies are required to clarify this point.

Anyway, the occurrence of an equilibrium ordered
phase appears to be clear from our numerical data, and is
in sharp contrast to
the vortex-glass (gauge-glass) 
problem where the screening was found to 
destabilize the equilibrium ordered phase.$^3$ Presumably, such 
difference comes from the fact
that the broken symmetry is a discrete $Z_2$ symmetry here
while
it is a continuous $U(1)$ symmetry in Ref.3.

We  also study the behavior of the flux.
The temperature dependence of 
the root-mean square of the local-flux amplitude, $\bar f$, 
defined by Eq.(3.10), is shown in Fig.4.
As can be seen from the figure, 
the magnitude of the induced local flux
is of order 0.1 flux quantum per plaquette for this inductance 
($\tilde {\cal L}=1$). 
Note that, in the small inductance limit $\tilde 
{\cal L}\rightarrow 0$,
$\bar f$ tends to zero, while  in the large 
inductance limit $\tilde {\cal L}\rightarrow \infty $,
$\bar f$ tends to 1/2 in the ground state 
of an isolated {\em frustrated\/} plaquette.

The flux Binder ratio, $g_{{\rm FG}}$,  defined by Eq.(3.7),
and the reduced flux-glass susceptibility, $\tilde \chi_{{\rm FG}}$,   
defined by Eq.(3.9), 
are shown in Fig.5,and 6, respectively.
Naively, one expects that the flux
should behave in the same way
as the chirality, since the
flux is also a pseudoscalar variable sharing the same symmetry 
as the chirality.
Indeed, as can be seen from Fig.6, the flux-glass susceptibility,
$\tilde \chi _{{\rm FG}}$, shows a  
divergent behavior similar to
$\tilde \chi_{{\rm CG}}$.
However, in contrast to the naive expectation, 
clear crossing of the Binder ratio 
as observed in $g_{{\rm CG}}$ 
is not observed in  $g_{{\rm FG}}$ at least 
in the range of lattice sizes studied here: see Fig.5. 
Rather, the ordering tendency seems 
more enhanced in the sense that
$g_{{\rm FG}}$ tends to increase with increasing $L$
exhibiting a feature of the ordered phase
even above 
$T_{{\rm CG}}\simeq 0.286 $. 
As the flux is an
{\em induced\/} quantity generated by the finite inductance effect, 
we believe this behavior to be a finite-size effect.
% caused by
Presumably, for this inductance value,
the flux hardly reaches its asymptotic critical behavior 
in rather small lattices studied here.
Recall  that, in the $\tilde {\cal L}\rightarrow 0$ limit, 
the flux vanishes trivially 
while the chiral-glass transition itself is 
most favored.
In fact, for larger inductances,
we have found that the Binder ratios of the flux and of
the chirality 
show more similar behavior as expected (see below).

We also compute the
zero-field linear and nonlinear susceptibilities,
$\chi $ and $\chi _2$, 
defined by Eqs.(2.9) and (2.10), respectively.
As can be seen from Fig.7, the linear
susceptibility
turns out to be paramagnetic, $\chi >0$, 
over an entire temperature range studied,
including in the disordered phase $T>T_{{\rm CG}}$, without a
clear anomaly at $T_{{\rm CG}}$. 
In shorter simulations on the same model
where the full equilibration is not achieved, 
$\chi $ tends to get smaller and sometimes becomes negative.$^{19}$ 
These results seem consistent with an earlier finding of 
Dom\'inguez {\em et al\/} who observed a paramagnetic $\chi $ in the
field-cooling mode, but a diamagnetic $\chi $ 
in the zero-field-cooling mode.$^{18}$
Meanwhile, the simulation of Ref.18
was performed for a rather large inductance, $\tilde {\cal L}=8$,
where the chiral-glass transition probably did not occur
in equilibrium (see below).
It should be stressed here that
the sign of $\chi $ is in fact a nonuniversal property:
Effects not taken into the present model, such as the
contribution of intragranular
supercurrents, could give additional diamagnetic contribution
in real systems and
could easily change the sign of the observed $\chi $.

By contrast, on general theoretical grounds, 
the nonlinear susceptibility, $\chi _2$, 
is expected to show a
negative divergence at the transition point where 
the time-reversal symmetry is spontaneously broken in a 
spatially random manner.$^{10}$ Indeed, as shown in Fig.8(a),
we have observed a behavior fully consistent with
this expectation. The exponent associated with this 
negative divergence 
is estimated via a finite-size scaling analysis with assuming
$T_{{\rm CG}}=0.286$ and $\nu _{{\rm CG}}$=1.3,  yielding 
$\gamma _2\simeq 4.4$ (see Fig.8b),
where $\chi _2\sim \mid T-T_{CG}\mid 
^{-\gamma _2}$.
This value of $\gamma _2$ is somewhat
larger than 
the chiral-glass susceptibility exponent, $\gamma _{{\rm CG}} 
\simeq 2.9$.
At present, it is not entirely clear
whether this  deviation reflects a true difference in the
asymptotic critical behavior. It appears likely that the
observed larger value of $\gamma _2$ simply comes from a finite-size 
effect related
to the possible nonasymptotic behavior of the flux 
as observed in $g_{{\rm FG}}$.
\bigskip\par

\noindent
[B] $\tilde {\cal L}=3,4$ and 5
\par\smallskip

So far, the results reported  were exclusively for the inductance 
$\tilde {\cal L}=1$.
We have also made similar, but less extensive simulations 
for other inductances 
$\tilde {\cal L}=3,4,5$ 
in order to
study the inductance dependence of the chiral-glass ordering.

In Figs.9 and 10, the temperature and inductance dependence of
the magnitude of the local chirality, $\bar\kappa $, and that of the
local flux, $\bar f$, are shown for a fixed lattice size $L=6$.
With increasing $\tilde {\cal L}$, 
$\bar\kappa $ tends to be suppressed
while $\bar f$ tends to be enhanced.
In Fig.11(a)-(c), the temperature dependence of 
the chiral Binder ratio,
$g_{{\rm CG}}$,  is shown for $\tilde {\cal L}=3,4$ and 5, respectively.
For $\tilde {\cal L}=3,4$, the curves of $g_{{\rm CG}}$ for
different $L$ appear to
cross at a finite temperature.
As expected, the chiral-glass
transition temperature monotonically decreases as 
$\tilde {\cal L}$ increases.
For $\tilde {\cal L}=5$, on the other hand, 
no crossing of $g_{{\rm CG}}$ is 
observed in the temperature range $T\geq 0.1$, 
suggesting that
the chiral-glass transition is highly suppressed at this inductance. 
We have also tried similar simulations for even larger inductances,
$\tilde {\cal L}>5$. However, the relaxation becomes extremely slow
for these larger inductances, and we can no longer 
equilibrate the system down to the temperature
range of interest  within a reasonable
amount of computation time.

The tendency that the chiral-glass ordering is suppressed at 
larger inductances can also be seen from other quantities, such as
the reduced chiral-glass susceptibility, $\tilde \chi _{{\rm CG}}$, 
the reduced flux-glass susceptibility, $\tilde \chi _{{\rm FG}}$, and
the nonlinear susceptibility, $\tilde \chi _2$.  
Thus, we show in Fig.12, 13  and 
14 the temperature and inductance dependence of 
$\tilde \chi _{{\rm CG}}$, $\tilde \chi _{{\rm FG}}$ and 
$\tilde \chi _2$
for a fixed lattice size 
$L=6$. As can clearly be seen in these figures, 
the chiral-glass ordering is more and more suppressed
for larger $\tilde {\cal L}$. By contrast, 
the paramagnetic tendency of
the linear susceptibility 
tends to be enhanced  for larger $\tilde {\cal L}$.
This is evident from Fig.15 in which the temperature
and inductance 
dependence of the linear susceptibility, $\chi $, is shown 
for a fixed lattice size $L=6$.

The obtained phase diagram in the $T-\tilde {\cal L}$ 
plane is sketched in Fig.16.
There appears to be a finite critical value of the inductance,
$\tilde {\cal L}_c$,
above which there is no equilibrium chiral-glass transition. 
Although
it is difficult to precisely locate $\tilde {\cal L}_c$
due to the
extremely slow relaxations we observed at lower temperatures,
it appears to lie around $5\lsim \tilde {\cal L}_c
\lsim 7$. If this is the case, the value of the inductance chosen
by Domingu\'ez {\em et al\/}$^{18}$ 
lied in the region of the
phase diagram where
no equilibrium chiral-glass transition took place. 
Then,  a kind of cooperative phenomenon accompanied with
a sharp growth of the
paramagnetic $\chi $, which was observed 
around $T\simeq 0.4$ in Ref.18, 
might not be related to an equilibrium phase transition, 
but be purely of dynamical origin.
This is consistent with our observation in Fig.15 
that the paramagnetic tendency is more enhanced 
for larger $\tilde {\cal L}$, while the chiral-glass ordering 
itself is suppressed
for larger $\tilde {\cal L}$.

In Figs.17(a) and (b), 
we show the temperature dependence of the flux Binder
ratio, $g_{{\rm FG}}$, for the case of $\tilde {\cal L}=4$ and 
$5$, respectively. In the case of $\tilde {\cal L}=4$, 
the curves 
of $g_{{\rm FG}}$ of $L=4$ and $L=6$
almost cross,
while they do not cross for $\tilde {\cal L}=5$.
Such behavior is more or less similar to the one observed in the 
corresponding chiral Binder ratio, $g_{{\rm CG}}$, for these
inductances: see Figs.11(b) and (c). This observation suggests that
for these  larger inductances 
the flux behaves in the same way as the chirality
even in rather small lattices
studied here, in contrast to the $\tilde {\cal L}=1$ case. 
\par
\bigskip
\bigskip
\setcounter{equation}{0}
\renewcommand{\theequation}{5.\arabic{equation}}
\noindent
\section {Summary and discussion}
\medskip

We have shown by extensive Monte Carlo simulations that
an equilibrium zero-field phase with a spontaneous broken
time-reversal symmetry, a chiral-glass phase, is possible
in certain ceramic superconductors with anisotropic
pairing symmetry. This phase is truly stable 
even in the presence of screening.
As in spin glasses, the nonlinear susceptibility exhibits
a negative divergence at the chiral-glass transition point. 
Via a finite-size scaling, 
static exponents associated with the chiral-glass
transition are determined. The obtained exponents 
are rather close to those of the 3D Ising spin glass. 
A rough phase diagram is constructed in the temperature-inductance
plane. It is found that the chiral-glass transition tends
to be suppressed for larger inductances, and there appears to be
a critical value of the parameter $\tilde {\cal L}$
beyond which there is no
equilibrium chiral-glass phase. By contrast,  
the paramagnetic tendency of the linear susceptibility (paramagnetic
Meissner effect) tends to be enhanced for larger inductances.
This observation clearly shows that, 
although the paramagnetic Meissner effect could also arise from the
frustration effect associated with the anisotropic nature
of the superconducting order parameter,
it has no direct relevance to the chiral-glass phase and the 
chiral-glass transition.

Next, on the basis of our findings in the present paper,
we wish to discuss  some requirements for the appropriate
ceramic samples where one could expect the chiral-glass phase.
One important parameter characterizing the sample is
the dimensionless inductance, $\tilde {\cal L}$, given by Eq.(2.3).
Our present result suggests that
an equilibrium chiral-glass state could be realized in the type of
samples with
smaller $\tilde {\cal L}$, 
but not for the samples with larger $\tilde {\cal L}$.
If one models a loop as 
a cylinder of radius $r$ and height $h$, its inductance is given by
${\cal L}=4\pi ^2r^2/h$.
Putting $r\sim 1\mu $m,
$h/r\sim 0.01$ and $J\sim 20$K 
(these values are chosen to mimic the sample used in Ref.16),
one gets $\tilde {\cal L}\sim 10^{-2}$.
Since this value is considerably smaller than $\tilde {\cal L}_c$,
an equilibrium chiral-glass phase may well occur
in such samples. By contrast, if the sample
has too large a grain size or too strong Josephson coupling,
an equilibrium chiral-glass phase will  not be realized, or at
least largely
suppressed.
Another requirement for the sample is
that the grains must be connected via weak links into an
infinite cluster,
not decomposed into finite clusters. Obviously,
finite-cluster 
samples cannot exhibit a chiral-glass transition, although
the paramagnetic Meissner effect is still possible.$^{6,9}$

Once appropriate samples could be prepared, the chiral-glass 
transition is detectable in principle via the standard
magnetic measurements by looking for a negative divergence of
$\chi _2$ or a magnetic ageing phenomenon, as in the case of
spin glasses.
In such magnetic measurements, care has to be taken to keep the 
intensity of applied ac and dc fields weak enough, typically much
less than 1G, so that the external flux per loop is sufficiently
smaller than $\phi _0$. 
Recently, a sharp negatively-divergent 
anomaly of $\chi _2$ was reported in a YB$_2$C$_4$O$_8$ ceramic
sample by the ac method by Matsuura {\em et al\/},$^{16}$ 
which might be a signal of the chiral-glass
transition. Ageing was observed in certain ceramic 
samples,$^{25}$ but not
in other samples.$^{9}$

As in the case of spin glasses,  measurements
of dynamic susceptibilities such as $\chi' (\omega )$ and
$\chi'' (\omega )$ would also give useful information on the 
possible chiral-glass ordering, particularly
when combined with the dynamic scaling analysis. For example,
near the  chiral-glass transition point, 
the imaginary part of the linear susceptibility, 
$\chi'' (\omega )$, is expected to satisfy
the  dynamic scaling relation of the form,
\begin{eqnarray}
\chi'' (\omega ,T,H)\approx \omega ^{ \beta _{{\rm CG}}
/z_{{\rm CG}}\nu_{{\rm CG}} } \bar \chi'' (
\frac { \omega }{t^{z_{{\rm CG}}\nu _{{\rm CG}} } },\ 
\frac {H^2 }
{t^{ \beta _{{\rm CG}} + \gamma _{{\rm CG}} } } ),)
\end{eqnarray}
where $t\equiv \mid (T-T_{{\rm CG}})/T_{{\rm CG}}\mid $ 
and $z_{{\rm CG}}$  is a dynamical chiral-glass exponent. From 
the present calculation, we get the static chiral-glass
exponents to be $\nu _{{\rm CG}}\simeq 1.3$, 
$\beta _{{\rm CG}}\simeq 0.5$ and $\beta _{{\rm CG}}+
\gamma _{{\rm CG}}\simeq 3.4$.
Although we cannot give a direct numerical estimate of
the dynamical exponent
from the present simulation, one might
guess that $z_{{\rm CG}}$ would
take a value around $z_{{\rm CG}}\nu _{{\rm CG}}
\simeq 7-8$ if one would assume the analogy
between the chiral glass and the  Ising spin glass
also in the dynamics.$^{26}$

Indeed,  a dynamic scaling analysis was made by 
Leylekian, Ocio and Hammann
for LSCO ceramic samples.$^{15}$ These authors performed 
both the ac susceptibility and the noise measurements, and 
found an
intergranular cooperative transition  even in
zero field at a
temperature about 10\% below the superconducting
transition temperature of the grain. 
Note that the noise measurements enable one to
probe truly zero-field phenomena where
one can be free from the extrinsic pinning effects 
such as the ones envisaged in the so-called critical-state
model.$^{27}$ 
It was then found that
the data of $\chi''$ satisfied the
dynamic scaling of the form (5.1).  
Here note that one is {\em not\/} allowed to invoke 
the standard vortex-glass scenario to
explain such intergranular cooperative transition 
{\em in zero field\/}, since in the standard vortex-glass picture
frustration is possible only under finite external fields.
By contrast, the experiment seems 
consistent with the chiral-glass picture.

Meanwhile, when the intragranular superconducting transition 
and the
intergranular transition  take 
place at mutually close
temperatures as in Ref.15, 
the Josephson coupling, $J$, which has been assumed
to be temperature independent in the present model, is 
actually strongly temperature dependent
in the transition region. In such a case, care has to be taken
in analyzing the experimental data, since such temperature
dependence of $J$ might modify the apparent 
exponent value from the true asymptotic value
to some {\em effective value\/}. 
In fact, the dynamical exponent $z\nu \simeq 30$ 
determined by Leylekian {\em 
et al\/} were  different from the 
standard spin-glass value,
which might be due to
the proximity effect of the intragranular
superconducting transition.$^{15}$
If, on the other hand, 
the temperature dependence of $J$ was taken into account 
in the fit, a
more realistic value $z\nu \simeq 10-15$ was obtained in Ref.15.
By contrast,
when the
intergranular chiral-glass transition takes place at a temperature
much lower than the intragranular superconducting transition 
temperature as in Ref.16,
the Josephson coupling can safely be regarded as
temperature independent as assumed in the present model.
Anyway, it is desirable to get a direct numerical 
estimate of the dynamical
chiral-glass exponent, $z_{{\rm CG}}$, to be compared with 
experiments. 
We are now planning to perform a  simulation to get an
independent numerical estimate
the dynamical exponent.

It may also be possible to detect a spontaneously induced flux in the
chiral-glass state by muon spin relaxation or electron holography
in zero external field.
As in the noise measurements, 
these measurements can be made in zero external field, 
and has 
an advantage of being free from the pinning effects of extrinsic 
origin. Here
it is essential to make such measurements for {\em ceramic\/} samples
with sufficiently many weak links, not for single crystals, simply
because the chiral-glass phase is expected only in the former
type of samples. By contrast, the kind of time-reversal-symmetry
breaking  proposed by Wen, Wilczek and Zee
is associated with the 
time-reversal-symmetry of the {\em bulk\/} 
superconducting order parameter
and should occur {\em even in single crystals\/}.$^{28}$

We could estimate an order of the induced flux 
below $T_{{\rm CG}}$ from the results of
our present simulation. In the case of
$\tilde {\cal L}=1$, for example, the flux intensity can be
estimated from  the calculated $[<q_f^2>]$ to be
about 0.02$\phi _0$ at 20\% below $T_{{\rm CG}}$.
For a sample with
a typical grain diameter about 1$\mu m$,
this corresponds to the flux intensity equal to 0.4G, 
which seems  well within
the sensitivity of the $\mu $sR measurements.
For a sample with a grain diameter about 10$\mu m$,
the flux intensity would be reduced to 4mG. 
If the dimensionless
inductance, $\tilde {\cal L}$, is significantly
smaller than unity, the flux intensity would become much
smaller, and eventually vanishes in the $\tilde
{\cal L}\rightarrow 0$ limit.

In the chiral-glass state,
the $U(1)$ gauge symmetry will {\em not\/} be broken,
even randomly, in the
strict sense. This means that the  
phase of the condensate, $\theta $, remains
disordered in the chiral-glass state at least on sufficiently
long length and time scales. Thus, 
the chiral-glass state
should not be a true superconductor, with a small but nonvanishing
linear resistance even below $T_{{\rm CG}}$.  
This property has been established in the
$\tilde {\cal L}\rightarrow 0$ limit where the screening effect can be
neglected.$^{10,17}$ Although  
we have not measured 
in the present simulation the quantity directly relevant to
the $U(1)$ gauge-symmetry breaking, 
the screening effect makes the interaction
between vortices short-ranged and makes the 
the $U(1)$ gauge-symmetry  breaking transition even more unlikely.

Small but finite linear
resistivity, $\rho _L$,  in the chiral-glass state
arises due to the
slow motion of thermally-activated {\em integer\/} vortex lines
(unbound vortex loops).
Free motion of integer vortex lines is still possible
in the chiral-glass state where chiralities (half-vortices) 
sitting at
frustrated plaquettes are frozen.
A schematic picture showing such free motion of integer vortex-line
excitations in the background of a frozen pattern of 
chiralities is given in Fig.18. One can see that 
free motion of integer-vortex lines of
either sign is possible without seriously destroying the 
freezing pattern of chiralities in the background.
In order to destroy the chiral-glass ordering in the background,
a chiral domain-wall-type excitation is necessary, which would
be responsible for the chiral-glass transition at $T=T_{{\rm CG}}$.
On decreasing the temperature across $T_{{\rm CG}}$, a sharp drop of
the resistivity will be caused by such chiral domain-wall excitations,
but the resistivity will stay finite even below  $T_{{\rm CG}}$ due to
the wandering vortex-line excitations.

We try to get a very rough order estimate of
$\rho _L$ at the chiral-glass transition point 
based on a  flux-creep model.$^{29}$ 
Within this model, the linear $I-V\/$ 
relation with finite $\rho _L$  is expected below a 
characteristic current-density scale, $j_c$, given by
$j_c\simeq k_BT/(\phi _0d^2\tilde \xi^2)$, where $d$ is a typical
grain size and $\tilde \xi$ is the phase 
correlation length (or the `spin' correlation length) in units of
$d$. We estimate $\tilde \xi$ at $T=T_{{\rm CG}}$ from the
Monte Carlo data of the 3D {\em XY\/} spin glass$^{30,17}$ to be
$\tilde \xi \sim 10$ lattice spacings. Here note that the spin-glass
correlation length does not diverge at the chiral-glass transition
point.$^{17}$ Then, 
for a typical ceramic sample, we put
$d\simeq 1\mu m$, $T=T_{{\rm CG}}\simeq 30K$, to get
$j_c\simeq 2\times 10^3$A/m$^2$.
The flux-creep model also yields 
$\rho _L\sim \phi_0^2d\tilde \xi/(k_BT\tau _0\tilde \tau)$, where
$\tau_0$ is an inverse `attempt frequency' of the intergranular vortex
motion, and $\tilde \tau$ is the phase or `spin' 
correlation time in units of
$\tau _0$. Again, from the Monte Carlo data,$^{30,17}$
we estimate
$\tilde \tau $ at $T=T_{{\rm CG}}$ to be $\tilde \tau 
\simeq 5\times 10^4$ Monte Carlo
time steps.
Precise value of our time unit, $\tau _0$, 
is largely unknown, but it should be
much longer than the atomic time scale since the vortex motion of
interest here is the one over grains. If we put
$\tau _0\simeq 10^{-9}$ sec, for example, we have 
$\rho _L\simeq 0.2\/\mu \Omega\cdot $cm  at $T=T_{{\rm CG}}$, 
while for 
$\tau _0\simeq 10^{-5}$ sec, we have 
$\rho _L\simeq 0.2\times 10^{-4}\/\mu \Omega\cdot $cm.
These values, though small, may be
within the reach of careful experimental measurements. 
\par

All simulations presented in this paper were done in zero 
external field.
A chiral-glass phase and a 
chiral-glass transition are associated with
a spontaneously breaking of time-reversal symmetry, and 
in that sense, can be regarded as a
zero-field phenomenon. Still, 
it should be emphasized here that the fate
of the chiral-glass phase and the chiral-glass transition in an
external field is not necessarily trivial and is of great
interest. Clearly, under external magnetic fields, 
the system no longer
possesses a global time-reversal symmetry.
Therefore, there cannot be a chiral-glass
transition associated with a spontaneous breaking of a 
{\em global\/} time-reversal symmetry.
Nevertheless, an interesting possibility emerges if the
chiral-glass transition in zero field accompanies the 
replica symmetry breaking$^{31}$ of the chirality. In such a case, 
an equilibrium phase and the associated thermodynamic transition
should persist even under finite fields and
are characterized
by the {\em chiral replica symmetry breaking\/}.
Then, the transition line in the $H-T$ plane might look like
the so-called AT-line familiar in spin glasses.$^{32}$
It is very interesting to relate such {\em chiral\/}-AT line
to the AT-like line often observed experimentally
in ceramic high-$T_c$ superconductors.$^{12}$ 

We finally note that the chiral-glass state can be realized not only 
in high-$T_c$ superconductors, but also in other superconductors
with nontrivial pairing symmetry, such as in heavy fermion 
superconductors, or possibly, in some organic superconductors.
It would be interesting to experimentally
search for this novel phase in these materials,
since it is a new state of matter realized only in
anisotropic superconductors with unconventional pairing symmetry.

The numerical calculation was performed on the FACOM VPP500
at the supercomputer center, Institute of Solid State Physics,
University of Tokyo. A part of this work was made when one 
of the authors (M.S. Li) was in Kyoto Institute of Technology.
M.S. Li thanks
the Japan Society for Promotion of Science for the award of
a fellowship. He was also supported in part by the Polish KBN grant.
\par
\newpage
\noindent
{\bf References}
\smallskip\par\noindent
\fussy
\begin{description}
\item 1 G. Blatter, M.V. Feigel'man, V.B. Geshkenbein, A.I. Larkin
and V.M. Vinokur, Rev. Mod. Phys. {\bf 66} (1994) 1125.
\item 2 M.P.A. Fisher,
Phys. Rev. Lett. {\bf 62} (1989) 1415.
\item 3 H.S. Bokil and A.P. Young, Phys. Rev. Lett. {\bf 74}, 
(1995) 3021; C. Wengel and A.P. Young, 
Phys. Rev. {\bf B54}  (1996) R6869.
\item 4 See, for example, 
A. Mathai, Y. Gim, R.C. Black, A. Amar and F.C. Wellstood,
Phys. Rev. Lett. {\bf 74}  (l995) 4523; 
D.J. van Harlingen, 
Rev. Mod. Phys. {\bf 67}, 515 (1995). 
\item 5 F.V. Kusmartsev, Phys. Rev. Lett. {\bf 69}  (1992) 2268;
J. of Superconductivity {\bf 5}  (1992) 463.
\item 6 M. Sigrist and T.M. Rice, 
J. Phys. Soc. Jpn. {\bf 61}  (1992) 4283; Rev. Mod. Phys.
{\bf 67} (1995) 503.
\item 7 P. Svedlindh, K. Niskanen, P. Nordblad, L. Lundgren,
B. L\"onnberg and T. Lundstr\"om, 
Physica {\bf C162-164}  (1989) 1365.
\item 8 W. Braunisch, N. Knauf, G. Bauer, A. Kock, A. Becker,
B. Freitag, A. Gr\"utz, V. Kataev, S. Neuhausen, 
B. Roden, D. Khomskii, 
D. Wohlleben, J. Bock and E. Preisler,
Phys. Rev. {\bf B48}  (1993) 4030.
\item 9 J. Magnusson,
J.-O. Andersson,  M. Bj\"ornander, P. Nordblad and P. Svedlindh, 
Phys. Rev. {\bf B51}  (1995) 12776; 
J. Magnusson, M. Bj\"ornander, L. Pust, P. Svedlindh,  
P. Nordblad and T. Lundstr\"om,
Phys. Rev. {\bf B52}  (1995) 7675.
\item 10 H. Kawamura, J. Phys. Soc. Jpn. {\bf 64}  (1995) 711.
\item 11 Note that 
the same state was also called `orbital-glass phase'
in Ref.10. Since the term `orbital-glass' has often been used in
the literature simply
to describe the state exhibiting the paramagnetic Meissner effect,
however, 
we donot use this terminology 
here to represent the thermodynamic phase
of our interest, but rather, use the term `chiral-glass',
following the {\it XY\/} spin-glass terminology. In the chiral-glass
phase, the susceptibility could be either paramagnetic or
diamagnetic.
\item 12 K.A. M\"uller, M. Takashige and J.G. Bednorz,  
Phys. Rev. Lett. {\bf 58}  (l987) 1143. 
\item 13 Z. Koziol, Physica {\bf C159}  (1989) 281. 
\item 14 K. Park, J.J. Kim and J.C. Park, Solid State Comm. {\bf 71}
(1989) 743.
\item 15 L. Leylekian, M. Ocio and J. Hammann, 
Physica C{\bf 185-189} (1991) 2243;  
Physica B{\bf 194-196} (1994) 1865.  
\item 16 M. Matsuura, M. Kawachi, K. Miyoshi, M. Hagiwara and 
K. Koyama,  
J. Phys. Soc. Jpn. {\bf 64}  (1995) 4540.
\item 17 H. Kawamura, Phys. Rev. {\bf B51}  (1995) 12398;.
J. Phys. Soc. Jpn. {\bf 61}  (1992) 3062; H. Kawamura and M. Tanemura,
J. Phys. Soc. Jpn. {\bf 60}  (1991) 608.
\item 18 D. Dom\'inguez, E.A. Jagla and C.A. Balseiro,
Phys. Rev. Lett. {\bf 72}  (1994) 2773.
\item 19 H. Kawamura and M.S. Li, Phys. Rev. 
{\bf B53} (1996) 619.
\item 20 K. Hukushima and K. Nemoto, J. Phys. Soc. Jpn.
{\bf 65}  (1996) 1604.
\item 21 H. Kawamura and M.S. Li, to appear in Phys. Rev. Lett. 
(1997).
\item 22 N. Kawashima and A.P. Young, 
Phys. Rev. {\bf B53}  (1996) 484;
A.P. Young and N. Kawashima,  Int. J. Mod. Phys. C{\bf 7} (1996) 327.
%R.N. Bhatt and A.P. Young, Phys. Rev. Lett. {\bf 54},  924 (l985); 
\item 23 K. Hukushima, H. Takayama and K. Nemoto,
Int. J. Mod. Phys. C{\bf 7} (1996) 337.
\item 24 J. Villain, J. Phys. C{\bf 11} (1978) 745.
\item 25 C. Rossel, Y. Maeno and I. Morgenstern,
Phys. Rev. Lett. {\bf 62}  (l989) 681. 
\item 26 A.T. Ogielsky, Phys. Rev. {\bf B32} (1985) 7384.
%A.T. Ogielsky, and I. Morgenstern, 
%Phys. Rev. Lett. {\bf 54}   (l985) 928.
\item 27 C.P. Bean, Rev. Mod. Phys. {\bf 36}  (1964) 31.
\item 28 X.G. Wen, F. Wilczek and A. Zee, Phys. Rev. {\bf B39} 
(1989) 11413.
\item 29 P.W. Anderson and Y.B. Kim, Rev. Mod. Phys. {\bf 36} 
(1964) 39.
\item 30 S. Jain and A.P. Young: J. Phys. {\bf C19} (l986) 3913.
\item 31 G. Parisi, J. Phys. A{\bf 13} (1980) 1101.
\item 32 J.R.L. de Almeida and D.J. Thouless, J. Phys. A{\bf 11}
(1978) 983.
\end{description}

\newpage
\noindent
{\bf Figure captions}
\medskip\noindent
\newcounter{fig}
\begin{list}{Fig. \thefig}{\usecounter{fig}\labelwidth32pt}
\item The temperature and size dependence of the root-mean
square of the local-chirality amplitude, $\bar \kappa $, 
defined by Eq.(3.4),
for $\tilde {\cal L}=1$. 
\item (a) The temperature and size dependence of the Binder 
ratio of the chirality, $g_{{\rm CG}}$, for $\tilde {\cal L}=1$. 
Inset is a magnified view around the
transition temperature $T_{{\rm CG}}\simeq 0.286$. 
(b) Finite-size scaling plot
of $g_{{\rm CG}}$ with $T_{{\rm CG}}=0.286$ and $\nu _{{\rm CG}}=1.3$.
\item (a) The temperature and size dependence of the reduced 
chiral-glass susceptibility, $\tilde \chi_{{\rm CG}}$, 
for $\tilde {\cal L}=1$. 
(b) Finite-size scaling plot
of $\tilde \chi_{{\rm CG}}$ with $T_{{\rm CG}}=0.286$,
$\nu _{{\rm CG}}=1.3$ and $\eta _{{\rm CG}}=-0.2$.
\item The temperature and size dependence of the root-mean
square of the local-flux amplitude, $\bar f$, defined by Eq.(3.10),
for $\tilde {\cal L}=1$. 
\item The temperature and size dependence of the Binder 
ratio of the flux, $g_{{\rm FG}}$, for $\tilde {\cal L}=1$. 
\item The temperature and size dependence of the reduced 
flux-glass susceptibility, $\tilde \chi_{{\rm FG}}$, 
for $\tilde {\cal L}=1$. 
\item The temperature and size dependence of the zero-field
linear susceptibility, $\chi $, for $\tilde {\cal L}=1$. 
An arrow in the figure represents the location of the
chiral-glass transition point.
\item (a) The temperature and size dependence of 
the zero-field nonlinear susceptibility, $\tilde \chi _2$, for 
$\tilde {\cal L}=1$. 
(b) Finite-size scaling plot
of $\tilde \chi _2$  with $T_{{\rm CG}}=0.286$, $\nu 
_{{\rm CG}}=1.3$ and
$\gamma _2=4.4$, where $\tilde \chi _2\sim (T-T_{{\rm CG}})
^{-\gamma _2} $.
An arrow in the figure represents the location of the
chiral-glass transition point.
\item The temperature and inductance 
dependence of the root-mean
square of the local-chirality amplitude, $\bar \kappa $, 
defined by Eq.(3.4),
for a fixed lattice size $L=6$.
\item The temperature and inductance 
dependence of the root-mean
square of the local-flux amplitude, $\bar f$, defined by Eq.(3.10),
for a fixed lattice size $L=6$.
\item The temperature and size dependence of the Binder 
ratio of the chirality, $g_{{\rm CG}}$, for (a) $\tilde {\cal L}=3$, 
(b) $\tilde {\cal L}=4$ and (c) $\tilde {\cal L}=5$.  
\item The temperature and inductance 
dependence of the reduced chiral-glass susceptibility, $\tilde\chi
_{{\rm CG}}$,
for a fixed lattice size $L=6$.
\item The temperature and inductance 
dependence of the reduced flux-glass susceptibility, $\tilde\chi
_{{\rm FG}}$,
for a fixed lattice size $L=6$.
\item The temperature and inductance 
dependence of the zero-field nonlinear susceptibility, 
$\tilde \chi_2$, for a fixed lattice size $L=6$.
\item The temperature and inductance 
dependence of the zero-field linear susceptibility, 
$\chi $, 
for a fixed lattice size $L=6$.
\item A phase diagram in the $T$-$\tilde {\cal L}$ plane.
Renormalized inductance $\tilde {\cal L}$ is defined by Eq.(2.3).
The data point at $\tilde {\cal L}=0$ is taken from Ref.17.
\item The temperature and size dependence of the Binder 
ratio of the flux, $g_{{\rm FG}}$, for (a) $\tilde {\cal L}=4$
and (b) $\tilde {\cal L}=5$.
\item Two-dimensional segment of the lattice showing 
thermally-activated integer vortex lines with 
vorticity $\pm 1$, wandering in the background
of a frozen pattern of chiralities in the chiral-glass state. 
Plus (+) and minus ($-$) chirality can be viewed as  half-vortices 
with vorticity $\pm 1/2$  sitting at frustrated plaquettes,
while unfrustrated plaquettes are frozen into the 
zero-chirality ($0$) state. If one looks at 
a given frustrated plaquette frozen into  
the + chirality (or vorticity $+1/2$) state, for example, 
its vorticity 
occasionally becomes $+3/2$ or $-1/2$ when 
the thermally-activated integer vortex line of either sign, +1 
or -1, passes this plaquette.
Still, the long-time average of the vorticity at this plaquette is
equal to $+1/2$, showing that the free motion of integer vortex lines
is compatible with the long-range chiral-glass order.
\end{list}

\end{document}